\documentstyle[epsf]{article}
\newcommand{\bra}[1]{\left\langle #1 \right|}
\newcommand{\ket}[1]{\left| #1 \right\rangle}
\newcommand{\omegaI}{\omega_I}
\makeatletter
\def\m@thcombine#1#2{%
  \setbox0=\hbox{$#1$}
  \setbox1=\hbox{$#2$}
  \ifdim\wd0>\wd1
    \setbox0=\hbox to\wd1{\hss\box0\hss}
  \else
    \setbox1=\hbox to\wd0{\hss\box1\hss}
  \fi
  \mathop{\vcenter{
    \offinterlineskip\box0\box1}}}
\def\lesim{\m@thcombine<\sim}
\def\gesim{\m@thcombine>\sim}
\def\lessgtr{\m@thcombine<>}
\def\gtrless{\m@thcombine><}
\makeatother
\renewcommand{\paragraph}[1]{
  \vspace{3mm}
  \leftline{\underline{\large\gt\bf #1}}
  \vspace{1.5mm}
  \noindent
}
\begin{document}
\Large
\hfill KUNS 1385\\
\hskip9mm
\begin{center}
Onset of Rotational Damping \\
in Superdeformed Nuclei
\end{center}
\large

\vspace*{0.5cm}
\noindent
\begin{center}
K.~Yoshida and M.~Matsuo$^{*}$
\end{center}

\noindent
\begin{center}
{\it Department of Physics, Kyoto University, Kyoto 606-01, Japan. }

\noindent
{\it $^{*}$ Yukawa Institute for Theoretical Physics, Kyoto
University, 
Kyoto 606-01,
 Japan. }
\end{center}

\small
\begin{minipage}[t]{155mm}

{\bf Abstract}:We discuss damping of the collective rotational motion in 
$A\sim 150$ superdeformed nuclei by means of a shell model
combining the cranked Nilsson mean-filed
and the surface-delta two-body residual force.
It is shown that, because of the shell structure associated with the
superdeformed mean-field, onset energy of the
rotational damping becomes 
$E_x \sim 2-3 $ MeV
above yrast line, which is much higher than in normal deformed nuclei.
The mechanism of the shell structure effect is investigated
through detailed analysis of level densities in superdeformed nuclei.
It is predicted the onset of damping varies in different
supedeformed nuclei along with variation  in the 
single-particle structure at the Fermi surface.
\end{minipage}
\vspace*{0.5cm}
\large

\section{Introduction}

\hskip 12pt
In recent experiments,
up to around 20 different rotational bands are
observed  in a normally deformed rare-earth nucleus, and 
several superdeformed rotational bands are identified 
in a single superdeformed nucleus
\cite{Dagnall94}.
The observed  rotational bands 
usually 
lie in the region near the yrast line and 
are often interpreted in terms of independent particle
excitations in a rotating deformed mean-field.
There
are of course numerous excited levels at higher excitation energy, but 
these levels do not necessarily form rotational band structure
characterized by sequences of strong E2 transitions
since damping of the collective rotational motion is expected
 \cite{Lauritzen86}.
In  the highly excited region, 
say at 1 to 2 MeV above yrast line for normally deformed nuclei, 
the level density at a given spin is as large as $10^2$ to
$10^3 $ MeV$^{-1}$, and the corresponding level spacing becomes 
comparable with or smaller than the
size of matrix elements (order of 10 keV) of the residual
two-body  interaction. The residual interaction then becomes effective
to cause  mixing of many-particle many-hole
configurations in the rotating deformed mean-field potential.
Since  different configurations respond differently to the
Corioli force 
due to different single-particle alignments, 
 the configuration mixing results in loss of
collectivity or damping of the collective rotational motion.
Experimental data on the damping of the rotational motion
are accumulating for  normally deformed rare-earth nuclei 
from the analysis of the quasi-continuum part of 
gamma-ray spectra containing gamma-rays emitted from the highly
excited levels \cite{Herskind92,Dossing95}.
In particular, a  fluctuation analysis 
of the quasi-continuum points to the presence of about 30 rotational
bands in 
a nucleus, thus confirms occurrence  of the rotational
damping \cite{Herskind92}. 

The mean-filed models which have been very successful 
for rotational bands near yrast line have to be
extended in order to describe the rotational damping caused by the
configuration mixing effect due to 
the residual interaction. Such extensions can be
formulated by using the shell model diagonalization on the
basis of many-particle many-hole configurations associated with the
cranked mean-filed[5-7].
In fact, a realistic model which combines the 
cranked Nilsson single-particle basis 
and the surface-delta residual two-body force \cite{Matsuo96} 
gives results consistent with 
the data from the fluctuation analysis method for normally deformed
rare-earth nuclei. It predicts that the rotational damping sets in at
around 1 MeV above yrast line.

The rotational damping  in superdeformed 
nuclei is expected to differ significantly 
from that in normally deformed nuclei 
since the level density of
superdeformed states is affected by the shell structure in the
single-particle spectra\cite{Aberg88,Dudek88}, 
which is the characteristic feature specific to the
superdeformation. The shell-effect on the rotational damping 
is pointed out by an previous work \cite{Aberg92} for $^{152}$Dy, which  
however adopted  schematic residual interaction 
with constant matrix elements treated as parameters. 
Recently, the fluctuation analysis method has been applied  
to a superdeformed  nucleus $^{143}$Eu with 
a first decisive and quantitative data for the onset of
the rotational damping in a superdeformed nucleus \cite{Leoni95}. 
It extracted an effective number of 
superdeformed rotational bands  which is not very
different from the effective number of bands in normally
deformed nuclei. 
Since the
shell structure in superdeformed nuclei is not uniform
even in $A\sim 150$ region, systematic and quantitative analysis
is required in order to reveal the characteristics of the
rotational damping. 

The present paper analyzes systematically the rotational damping in 
superdeformed nuclei in the $A\sim 150$ region by means of
the shell model diagonalization with realistic two-body 
residual interaction \cite{Matsuo93}. In particular,
the analysis is focused on the onset of the damping since
it is most relevant to the data extracted from
the fluctuation analysis method. In order to clarify the
shell structure effect on the rotational damping, 
we also look into the level densities in detail in Section
\ref{leveldns}.
It is revealed that, as  consequence of the shell structure effects, 
the onset of damping varies in different species of 
superdeformed nuclei.
This is because the single-particle
spectra as well as the shell-gap at the Fermi surface
varies as $N$ and $Z$ change.


\section{Formulation}
\label{form}
\hskip 12pt
The microscopic model adopted in the present study 
is  essentially the same as the formulation 
for normally deformed nuclei \cite{Matsuo96}, except
difference in the mean-field potential. 
In describing superdeformed states, suitable choice
of the deformation parameters is important since the shell
structure characteristic to superdeformation 
depends on the deformation. 

We start with the cranked Nilsson single-particle Hamiltonian
\begin{equation}
\hat{h}=\hat{h}_{\rm Nilsson}-\omega \hat{j}_x
\label{nilham}
\end{equation}
with $\omega$ being the rotational frequency.
The pairing potential is not introduced. 
For the Nilsson potential, the quadrupole and hexadecapole
parameters $\epsilon_2$ and $\epsilon_4$ are considered
as shape variables. We adopt the Nilsson parameters
given by Ref.~\cite{Nilsson69}, which gives better description of
proton $7_1$ orbit than the parameter
\cite{Bengtsson85} used in Ref.~\cite{Matsuo93}. 
The difference in the Nilsson parameters does not
affect the following argument except detailed behavior.
With reflection symmetry,
the single-particle routhian orbits of the cranked Hamiltonian
keep the parity and the signature quantum numbers.
In order to guarantee smooth change in the single-particle
spectrum   as a function of the rotational frequency,
an diabatic basis for the single-particle orbits is constructed
in a similar way to Ref. \cite{Bengtsson89}.

\begin{figure}[t]
\centerline{
\epsfysize=10cm\epsffile{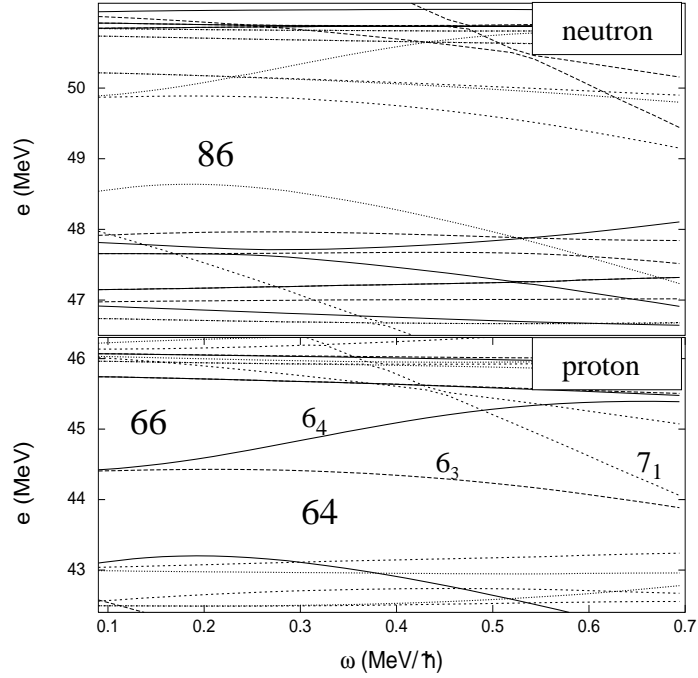}
}
\caption{\label{gdsp}The single-particle routhian spectra 
for $^{152}$Dy with the equilibrium deformation determined
by the energy minimization. The deformation parameters
$(\epsilon_2,\epsilon_4)$  change continuously from (0.577,0.010)
at $\omega=0.07$(MeV/$\hbar$),  $I=10 \hbar$ to (0.570,0.030) at
$\omega=0.74$(MeV/$\hbar$), $I=70 \hbar$ .
}
\end{figure}

An reference superdeformed configuration is defined for a given nucleus 
by putting $N$ neutrons and $Z$ protons up to the
Fermi surface in the diabatic single-particle
spectrum. When the reference configuration is not unique
due to 
crossings at the Fermi surface, we choose
one by referring to the high-N configuration suggested from
the observed yrast superdeformed band \cite{Bengtsson-Aberg89}.
Deformation parameters $(\epsilon_2$,$\epsilon_4)$ are
determined as functions of spin by minimizing the
 total energy of the reference configuration with 
the Strutinsky shell correction defined at a given spin
$I$\cite{Anderson76} 
\begin{equation}
E(I)=E^{\rm micro}(I)-E^{\rm smooth}(I)+E^{RLD}(I) .
\label{toteng}
\end{equation}
Here the microscopically calculated total energy is given by
\begin{equation}
E^{\rm micro}(I)=\sum_{
\small
\begin{array}{c}
{\rm occupied}\;i \\
\end{array}
}
e_i(\omegaI)+\omegaI I
\end{equation}
with the rotational frequency $\omegaI$ given
\begin{equation}
J_x(\omegaI)=\sum_{{\rm occupied}\; i} j_{xi}(\omegaI)=I,
\end{equation}
and $e_i(\omega)$ and $j_{xi}(\omega)$ being the diabatic
single-particle 
routhian energy and expectation value of $\hat{j}_x$. The Strutinsky
smoothed total energy is given by
\begin{equation}
E^{\rm smooth}(I)=\sum \tilde{e}_i(\tilde{\omega})+\tilde{\omega}I
\end{equation}
and
\begin{equation}
\tilde{J}_x(\tilde{\omega})=\sum \tilde{j}_{xi}(\tilde{\omega})=I,
\end{equation}
with $\tilde{e}_i,\tilde{j}_{xi}$ being $e_i,j_{xi}$
weighted by the smoothed occupation number $\tilde{n}_i$
\cite{RingSh}. \\
The rotating liquid-drop energy $E^{RLD}(I)$ is calculated according
to Ref.~\cite{Myers-Swiatecki67}.  
\begin{table}
\begin{center}
\begin{tabular}{|c|c|c|}
\hline
& $\epsilon_2$ & $\epsilon_4$ \\
\hline
$^{143}$Eu& 0.478 & 0.044 \\
$^{146}$Gd& 0.516 & 0.035 \\
$^{147}$Gd& 0.523 & 0.037 \\
$^{148}$Gd& 0.534 & 0.031 \\
$^{149}$Gd& 0.545 & 0.028 \\
$^{150}$Gd& 0.556 & 0.024 \\
$^{150}$Tb& 0.555 & 0.028 \\
$^{151}$Tb& 0.565 & 0.023 \\
$^{152}$Tb& 0.577 & 0.022 \\
$^{151}$Dy& 0.564 & 0.026 \\
$^{152}$Dy& 0.574 & 0.021 \\
$^{153}$Dy& 0.584 & 0.019 \\
\hline
\end{tabular}
\caption{\label{tabdef} 
The equilibrium deformation parameters $\epsilon_2$ and $\epsilon_4$
at $I=50\hbar$  determined by Strutinsky
method for each superdeformed nucleus. 
}\end{center}
\end{table}
The deformation parameters thus determined are listed in
Table.~\ref{tabdef} 
at a representative spin $I=50\hbar$. The spin dependence of the
deformation is weak in all the nuclei under consideration, {\it e.g.}
$(\epsilon_2,\epsilon_4) = (0.577,0.010)\sim (0.570 0.030)$ for
$I=10 \sim 70\hbar$ in $^{152}$Dy. 
An example of the calculated  cranked Nilsson  spectra with 
the optimized deformation  is shown in Fig.~\ref{gdsp} for $^{152}$Dy.

In a superdeformed nucleus,there are usually more than two coexisting 
local minima of the total energy surface in the deformation parameter
space, which correspond to superdeformed and weakly deformed
mean-fields. If the mean-fields are stable(the minima are deep),excited
states are expected to be built upon each mean-field. At high
spins ($I\gesim 40\hbar$) where the superdeformed minimum becomes the
lowest,  states near the yrast line are dominated by 
excited superdeformed states.

The basis for the excited superdefomed states is formed by
many-particle many-hole ($n$p-$n$h) configurations in the diabatic
cranked Nilsson orbits which cover the energy range 
$\delta E= \pm $4.8MeV above and below the Fermi surface. 
The deformation parameters and rotational frequency specifying the
single-particle basis for  
$n$p-$n$h configurations are taken common to those for the reference
configuration. 
Energy of the basis states is given by 
Eq.~(\ref{toteng}), in which the microscopic energy is
given for each $n$p-$n$h configuration(labeled by $\mu$) 
in the same way as in Ref.~\cite{Aberg88},{\it i.e.},
\begin{equation}
E_{\mu}^{\rm micro}(I)=E_{\mu}(\omegaI)+\omegaI I+\frac{1}{2 {\cal
J}_{\mu}^{\rm micro}}(I-J_{x\mu}(\omegaI))^2
\end{equation}
where 
\begin{equation}
E_{\mu}(\omegaI)=\sum_{
\small
\begin{array}{c}
{\rm occupied}\;i \\
{\rm in}\;\mu
\end{array}
}
e_i(\omegaI)
\end{equation}
and 
\begin{equation}
{\cal J}_{\mu}^{\rm micro}=
\sum_{
\small
\begin{array}{c}
{\rm occupied}\;i \\
{\rm in}\;\mu
\end{array}
}
\left.\frac{dj_{xi}(\omega)}{d\omega}\right|_{\omega=\omegaI}
\end{equation}
is the dynamic moment of inertia of the configuration $\mu$.
The Strutinsky renormalization is important not only to determine
the deformation but also to 
describe  gamma-ray energies of the collective E2 transitions.
The basis of the excited superdeformed states is denoted  as 
$\{\ket{\mu (I)}\}$ with the energy $\{E_\mu(I)\}$. 
It  keeps good quantum numbers 
with respect to the total signature and 
parity. Excited states built
upon weakly deformed mean-fields are neglected in the present paper.
At low spins ($I\lesim 40\hbar$), they may dominate the yrast structure and
the superdeformed states are to be surrounded by these weakly deformed 
states, while coupling between the superdeformed and weakly deformed
excited states is expected to be weak if there exists a potential barrier
between the local minima. We neglect the presence of
the weakly deformed states since the coupling is hardly treated by a
shell model description. Hereafter the yrast line 
refers to an envelope of the lowest energy superdeformed states as a
function of spin.

For the residual two-body interaction, it is natural to assume
that the effective nuclear force  is not dependent on
the deformation. We adopt the same surface-delta
interaction (SDI) \cite{Mozkowski65} with the same strength as used
for describing the rotational 
damping in normally deformed nuclei so that results here
can be compared with those for normally deformed nuclei
\cite{Matsuo96}.
The strength of SDI is 
 $V_0 =27.5/A$(MeV), which is  taken from Ref.~\cite{Faessler68}.
The two-body interaction is decomposed into mean-field part associated
with the reference configuration and
its residual, and only the residual two-body force is taken into
account. 
The residual interaction causes configuration mixing among the
basis states $\{\ket{\mu(I)}\}$ 
and generates energy eigenstates $\ket{\alpha(I)}$ 
of the many-body Hamiltonian,
\begin{equation}
\ket{\alpha(I)}=\sum_{\mu}X_{\mu}^{\alpha}(I)\ket{\mu(I)}
\end{equation}
whose amplitudes are determined by numerical diagonalization.
The diagonalization is done separately for each spin and parity with
the lowest 700 basis states. 

The E2 operator associated with the collective rotation
is given by
\begin{equation}
\bra{\mu(I)}M(E2,\lambda=2,\mu_x=2)\ket{\mu'(I\! -\! 2)}\\
=\sqrt{\frac{15}{128\pi}}Q_0\delta_{\mu\mu'},
\label{assume1}
\end{equation}
assuming that the collective E2 transition takes place between
states having the same microscopic configuration.
Static quadrupole moment $Q_0$ is assumed to be constant for all
configurations.
The E2  strength for the transition from a level $\alpha$ at $I$ to 
a level $\beta$ at $I\! -\! 2$ is given by
\begin{equation}
M^2_{\alpha I\beta I\! -\! 2}\\
=\frac{15}{128\pi}Q_0^2 w_{\alpha\rightarrow\beta}
\end{equation}
with normalized strength 
\begin{equation}
w_{\alpha\rightarrow\beta}=\left(\sum_{\mu}X_{\mu}^{\alpha
*}X_{\mu}^{\beta}\right)^2,
\label{eqbranch}
\end{equation}
\begin{equation}
\sum_{\beta}w_{\alpha\rightarrow\beta}=1 \ \ .
\end{equation}

\section{Level density}
\label{leveldns}
\begin{figure}
\begin{minipage}[tbp]{70mm}
\rightline{
\epsfysize=7.5cm\epsffile{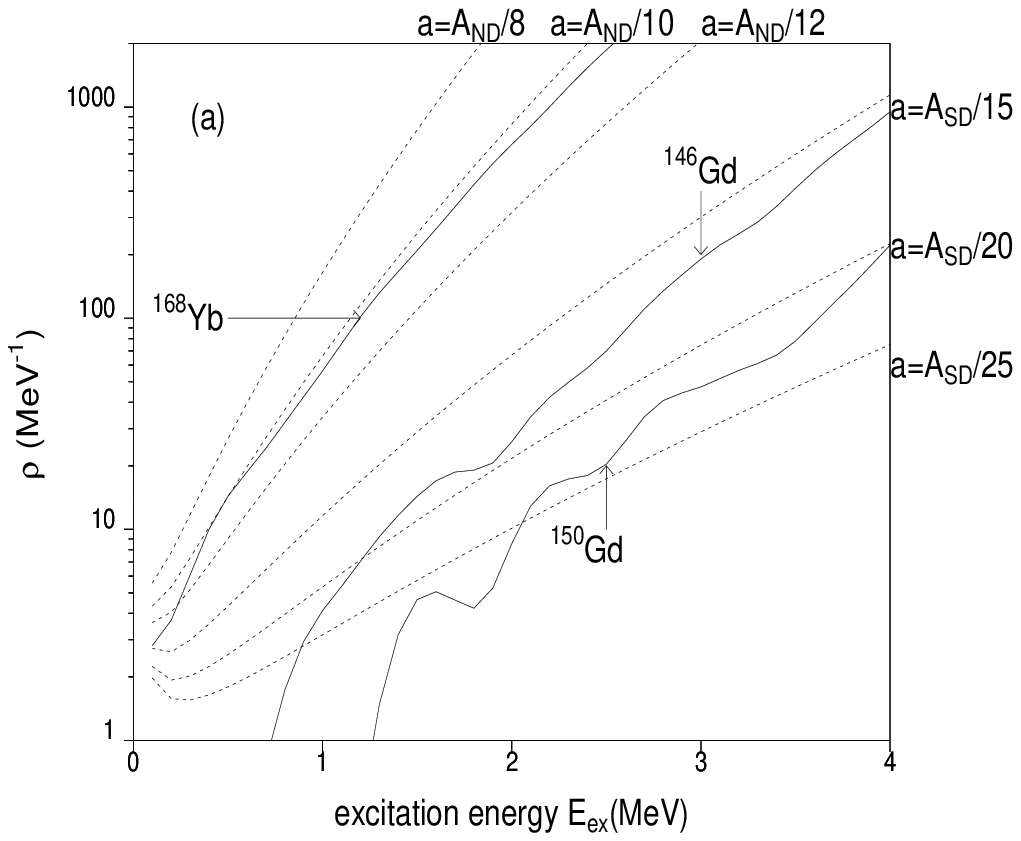}
}
\rightline{
\epsfysize=7.5cm\epsffile{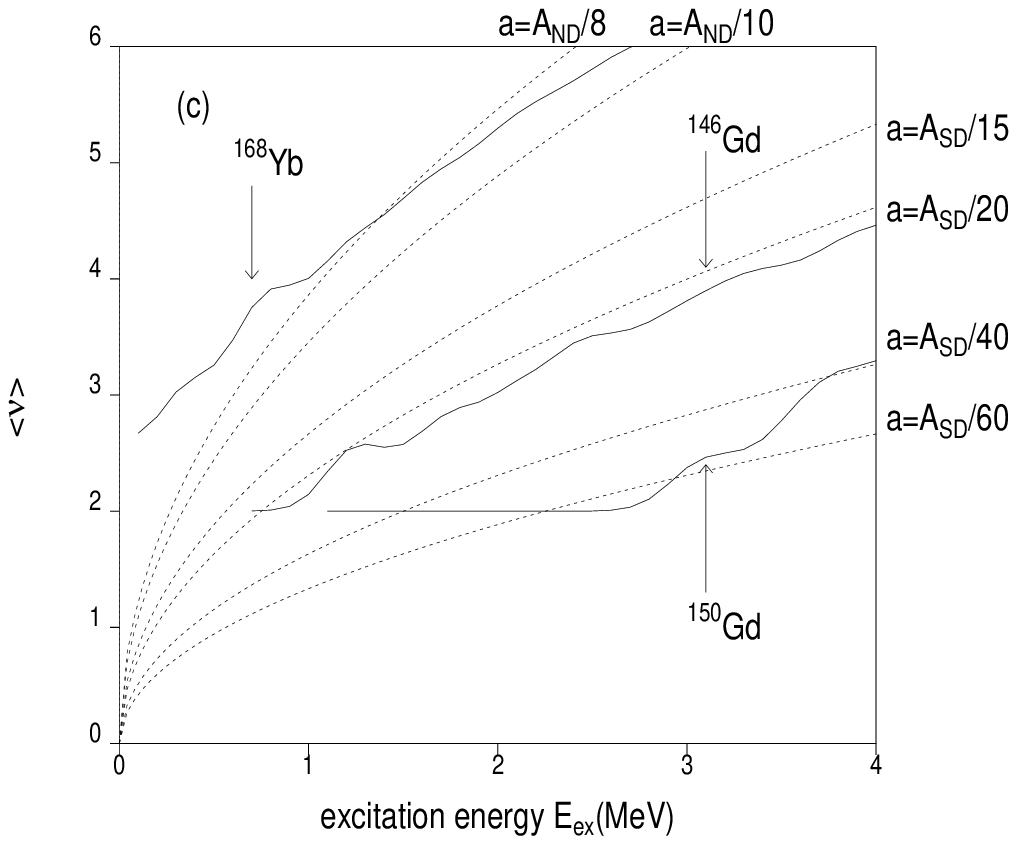}
}
\end{minipage}
\hskip 10mm
\begin{minipage}[t]{70mm}
\rightline{
\epsfysize=7.5cm\epsffile{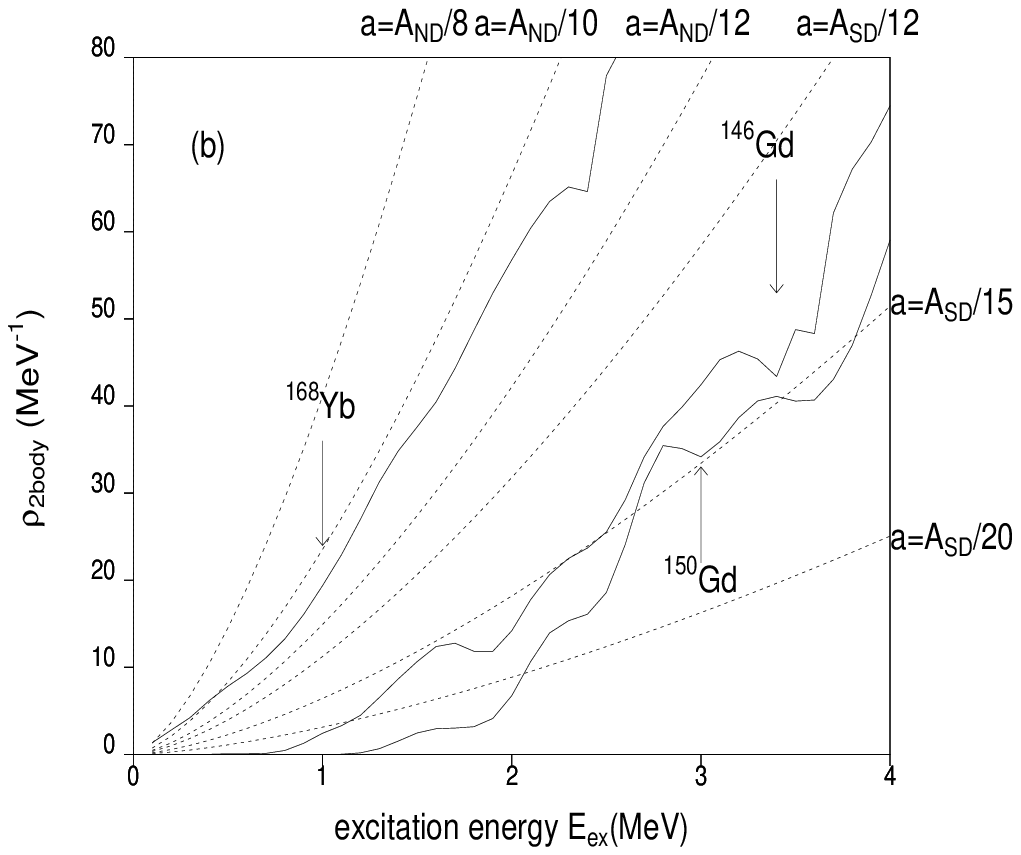}
}
\caption{\label{ferns}(a)Level densities $\rho(E)$,(b)two-body level
densities $\rho_{\rm 2body}(E)$ and (c) average seniorities
$\left\langle \nu \right\rangle (E)$ 
of superdeformed states as functions of the excitation energy
measured from the yrast line at $I=50\hbar$ in $^{146}$Gd,
and those for the normally deformed nucleus $^{168}$Yb at $I=40\hbar$.
The level density is averaged over both parities.
The Fermi gas expressions are also plotted with dotted curves
for various values of level density parameter $a$, in which $A_{\rm
SD}=150$ for superdeformed states and $A_{\rm ND}=168$ for ${}^{168}$Yb.}
\end{minipage}
\end{figure}

\hskip 12pt
In this section, we discuss  characteristics of level densities of
superdeformed nuclei. Since our purpose is to  understand
mechanism of the 
configuration mixing and the rotational damping, 
we deal with the level densities before
the shell model diagonalization, {\it i.e.}, all the calculations in this
section refer to the unperturbed basis states $\{\ket{\mu(I)}\}$ 
and their
energies $\{E_\mu(I)\}$. 
Fig.~\ref{ferns}(a) shows calculated level density of 
 superdeformed states and compares it with the level density
in  a typical normally deformed 
nucleus $^{168}$Yb. 
The calculated level density of super-deformed states is
systematically  much lower by an order of magnitude 
than that of normally deformed states. A similar results is obtained in
the previous analysis  \cite{Aberg88,Dudek88}.
This is attributed to the low 
level density of the single-particle levels
at the Fermi surface (See Fig.\ref{gdsp}),
which is an general feature  intrinsic to the
superdeformed single-particle potential.
In order to give quantitative discussion, 
the calculated level density is compared in Fig.~\ref{ferns}(a)
with the 
Fermi-gas expression\cite{Bohr-Mottelson69I,Aberg88}
\begin{equation}
\label{fermtot}
\rho(E)=\frac{\sqrt{\pi}}{48}a^{-\frac{1}{4}}
E^{-\frac{5}{4}}\exp{2\sqrt{aE}}
\end{equation}
where the level density parameter $a$ is supposed to be reduced and
is treated as a variable parameter.
The level density 
in the normal deformed nucleus $^{168}$Yb is well fitted by
the Fermi-gas expression with 
the level density parameter $a = \frac{\pi^2}{6} g_0\sim A/10$, 
which agrees with
the standard 
estimate of the single-particle level density $g_0$
\cite{Bohr-Mottelson69I}. 
In contrast, the calculated level density 
of superdeformed states 
corresponds to much smaller value around
$a=A/15\sim A/25$. It is known from fitting experimetal data to the
Fermi gas formula that the level density 
parameter in spherical closed shell nuclei is sizably smaller than the 
over-all systematics\cite{Bohr-Mottelson69I}.

Another level density function  which counts the states
connected by the two-body residual interaction 
plays an important role in
the configuration mixing of unperturbed states\cite{Lauritzen86}.
This quantity, which we call the two-body level density
$\rho_{\rm 2body}(E)$ in the following,
can be calculated explicitly as described in Appendix. 
The result is depicted in Fig.~\ref{ferns}(b) and is also compared
with the $\rho_{\rm 2body}$ in normally deformed
$^{168}$Yb and 
the Fermi-gas estimate for a fixed spin and parity\cite{Lauritzen86}
\begin{equation}
\label{twobody}
\rho_{\rm 2body}=\frac{81}{4 \pi^6}a^{5/2}E^{3/2}.
\end{equation}
The two-body level density for superdeformed states is
again much lower than that in the normally deformed nucleus.
The corresponding value of the level density parameter
is around  A/12 to A/20.

Fig.~\ref{ferns}(c) shows that average seniority(average number of
particles and holes excited from the reference 
configuration).This quantity for superdeformed states is also
significantly low in comparison with 
the normally deformed. The level density parameter
 $a_{\rm seniority}$ which is obtained by a fit to the 
Fermi gas
expression~\cite{Ericson60,Aberg88} 
\begin{equation}
\left\langle \nu \right\rangle=\log 4\left( \frac{36}{\pi^4}aE \right)
\end{equation}
is very low, ranging around A/15 to A/60.
The level density parameters  $a_{\rm tot}$,  $a_{\rm 2body}$
and  $a_{\rm seniority}$ 
obtained from the calculated 
$\rho(E)$, $\rho_{\rm 2body}(E)$, and $\left\langle \nu
\right\rangle(E)$
are  listed 
in Table.~\ref{tabdns}.
\begin{table}
\begin{center}
\begin{tabular}{|c|c|c|c|c|}
\hline
&$A/a_{\rm tot}$[MeV] & $A/a_{\rm 2body}$[MeV] & $A/a_{\rm
seniority}$[MeV] & $E_{\rm shell}$[MeV] \\  
\hline
$^{143}$Eu& 18.5 &        14.9 & 25.2    & 3.56  \\
$^{146}$Gd& 17.6 &        15.0 & 15.9    & 3.94  \\
$^{147}$Gd& 17.5 &        14.2 & 28.0    & 4.02  \\
$^{148}$Gd& 19.1 &        14.6 & 30.8    & 4.31  \\
$^{149}$Gd& 19.4 &        14.1 & 33.8    & 4.59  \\
$^{150}$Gd& 23.9 &        17.0 & 56.2    & 5.14  \\
$^{150}$Tb& 19.0 &        15.1 & 30.3    & 4.43  \\
$^{151}$Tb& 22.6 &        16.4 & 36.6    & 5.08  \\
$^{152}$Tb& 18.2 &        13.4 & 23.0    & 4.53  \\
$^{151}$Dy& 13.8 &        12.3 & 14.8    & 3.09  \\
$^{152}$Dy& 18.9 &        16.6 & 26.7    & 4.55  \\
$^{153}$Dy& 15.9 &        12.1 & 20.3    & 4.19  \\
\hline
\end{tabular}
\caption{\label{tabdns}The level density parameters
$a_{\rm tot},a_{\rm 2body}, a_{\rm seniority}$,which are obtained by a 
fit to the
calculated level density  $\rho(E)$, the two-body level
density $\rho_{\rm 2body}(E)$, and the average
seniority $\left\langle \nu \right\rangle(E)$ at  $E=2.5$ MeV and
$I=50\hbar$.  
Since the level density parameter is often expressed as
$a=A/a_0$, the table lists the value of $a_0$ in unit of MeV 
instead of $a$.
The fifth column lists the shell correction energy 
 $E_{\rm shell}=-(E^{\rm micro}-E^{\rm smooth})$.
}
\end{center}
\end{table}

It should be emphasized  that 
 the level densities $\rho(E)$ and $\rho_{\rm 2body}(E)$,
 and the average seniority  $\left\langle \nu \right\rangle(E)$ for
superdeformed states are not well fitted
by the Fermi-gas formulas with  a single value of $a$ 
for the superdeformed states. Apparently
the excitation energy dependence of the calculated level densities
deviates from the Fermi gas formula 
in the excitation energy range displayed in 
 Fig.~\ref{ferns}.
For example, the total density $\rho(E)$ increases
very slowly 
for $E<1$ MeV while it shows steeper increase  at
$E>1$ MeV  than the fitted Fermi gas formula. 
It is also noticed that
the level density parameter, say  $a_{\rm tot}$,   which fits the
calculated  
level density $\rho$ at a given excitation energy differs from
the parameter $a_{\rm 2body}$ fitting $\rho_{\rm 2body}$
and from $a_{\rm seniority}$; $a_{\rm tot}$ is systematically  smaller
than $a_{\rm 2body}$, and $a_{\rm seniority}$ being further smaller 
than $a_{\rm tot}$.
These features clearly indicate that the level densities of the
superdeformed states cannot be represented by the Fermi gas model.
The deviation between $a_{\rm tot}$ and $a_{\rm seniority}$ is
previously pointed out\cite{Aberg88}.
\begin{figure}[tbp]
\centerline{
\epsfysize=7.5cm\epsffile{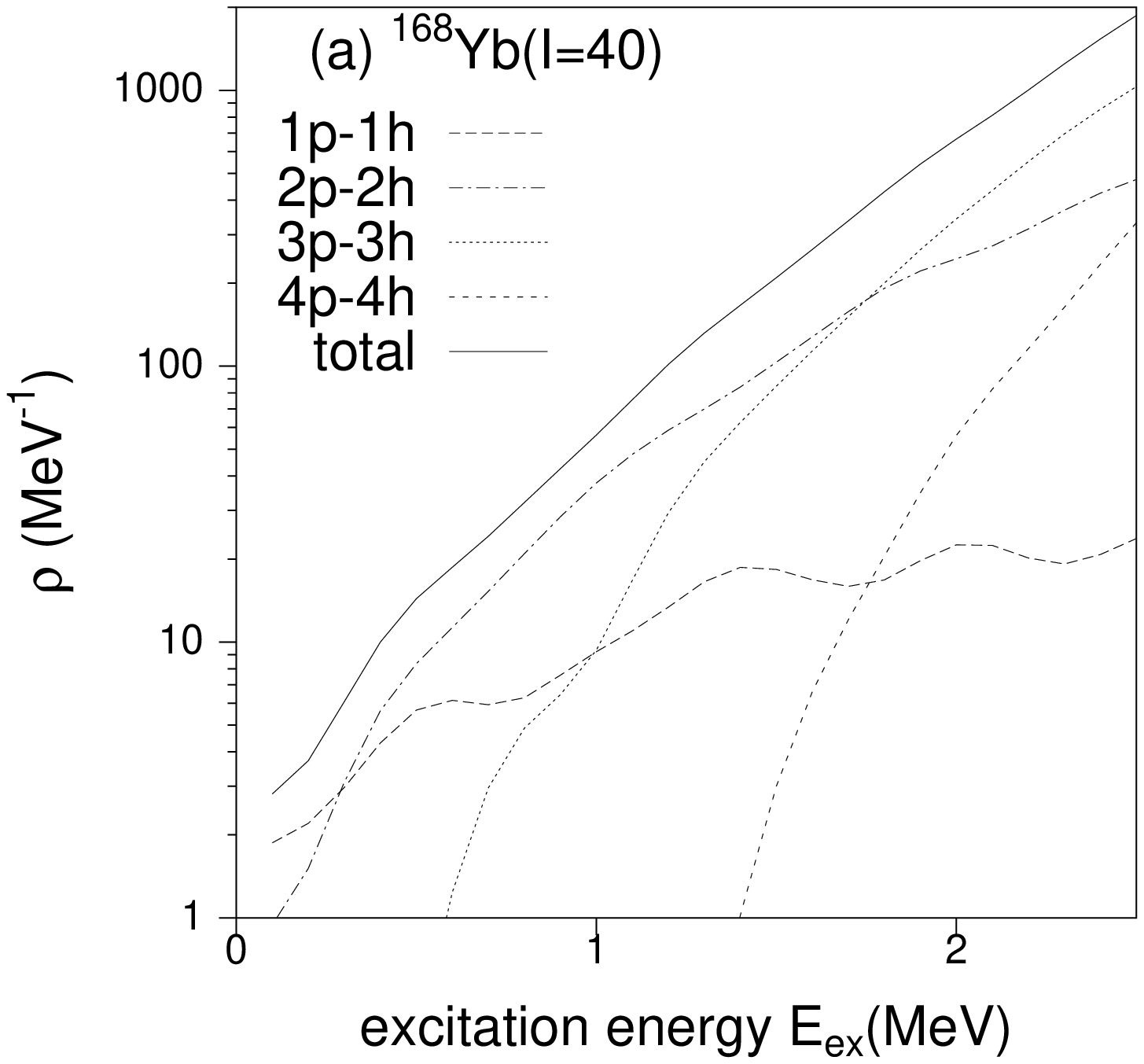}
\epsfysize=7.5cm\epsffile{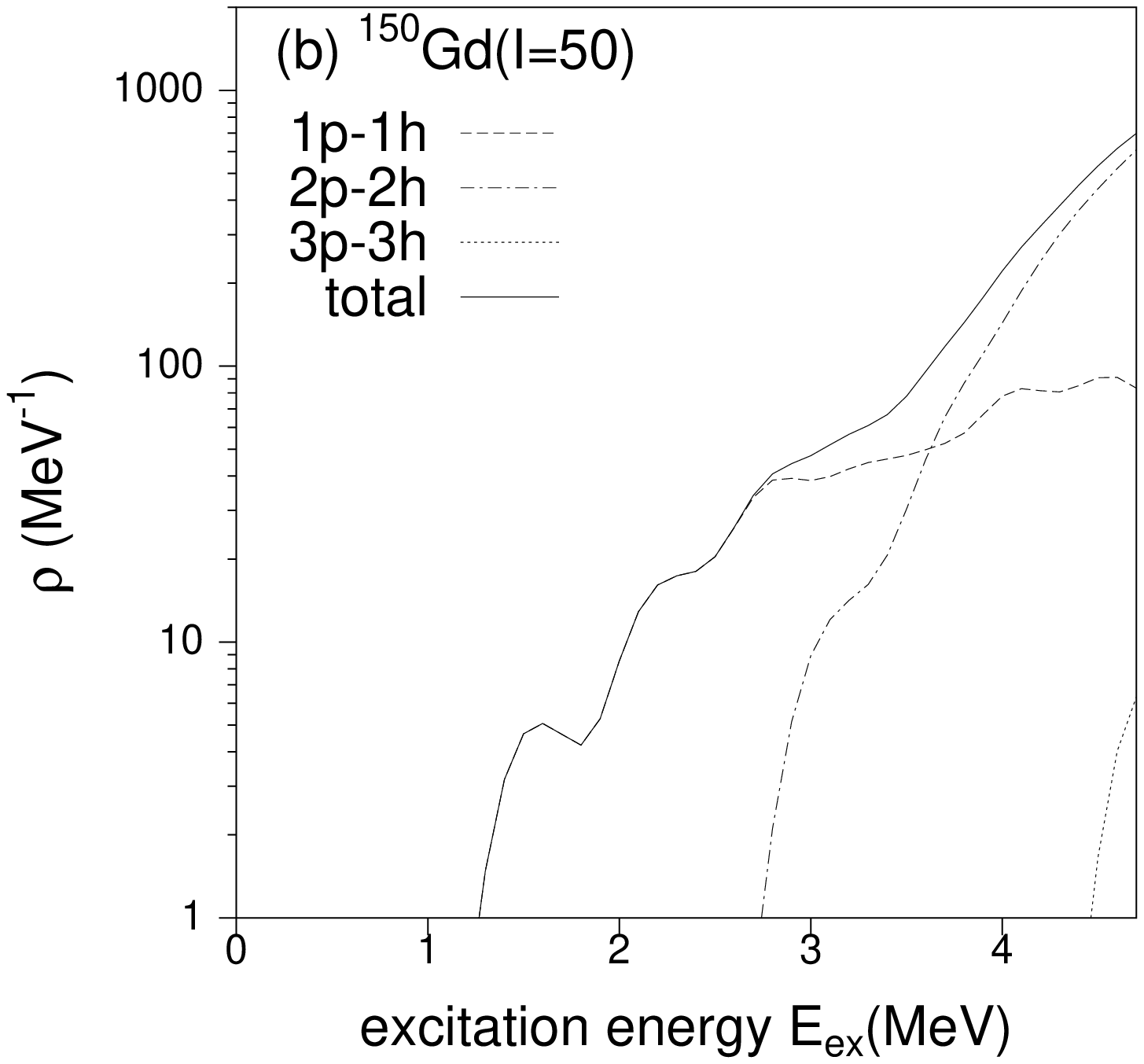}
}
\caption{\label{npnh}
The level density of unperturbed $n$p-$n$h configurations for
(a)normally deformed nucleus $^{168}$Yb at
$I\! =\! 40\hbar$ and (b)superdeformed $^{150}$Gd at $I\! =\!
50\hbar$ as a function of the excitation energy
measured from the yrast line.
In $^{150}$Gd only 1p-1h and 2p-2h components are dominant for $E_{\rm 
ex}<3$ MeV,while in $^{168}$Yb 1p-1h to 4p-4h contributes to the level
density at the same energy region. 
}
\end{figure}

In order to understand the deviation from the Fermi gas model,
it is useful to classify the excited states in terms
of the number of excited particles and holes. The decomposed
level densities are shown in Fig.~\ref{npnh}
for superdeformed states 
$^{150}$Gd as well as for 
normally deformed $^{168}$Yb.
In $^{168}$Yb, there are contributions from 1p-1h to 4p-4h states 
for excitation below 3 MeV, while only 1p-1h and 2p-2h states
contribute in the same energy region in superdeformed $^{150}$Gd.
Superdeformed states with 3p-3h and higher $n$p-$n$h configurations lie
at higher energy region. 
Note that
the shell structure in superdeformed nuclei produces the gap
of about 1 MeV at the Fermi surface in the single-particle spectrum , 
because of which each particle-hole
excitation costs at least about 1 MeV of excitation energy.
Such a feature cannot be represented by
the Fermi gas model which assumes
uniform single-particle spectrum. 

The difference 
between $a_{\rm tot}$ and $a_{\rm 2body}$ is also explained by the
presence of 
shell gap. 
The level density of $n$p-$n$h states 
depends on excitation energy  as $\rho_{npnh}\propto E^{2n-1}$ for an
equidistant 
single particle spectrum\cite{Ericson60} ,which is
modified as 
$\rho_{npnh} \propto (E-n\Delta)^{2n-1}$ by the
presence of the gap $\Delta$ at the Fermi surface.
The decrease  of the density due to the gap becomes very significant
for $n$p-$n$h states with large $n$.
\begin{figure}[tbp]
\centerline{
\epsfysize=22cm\epsffile{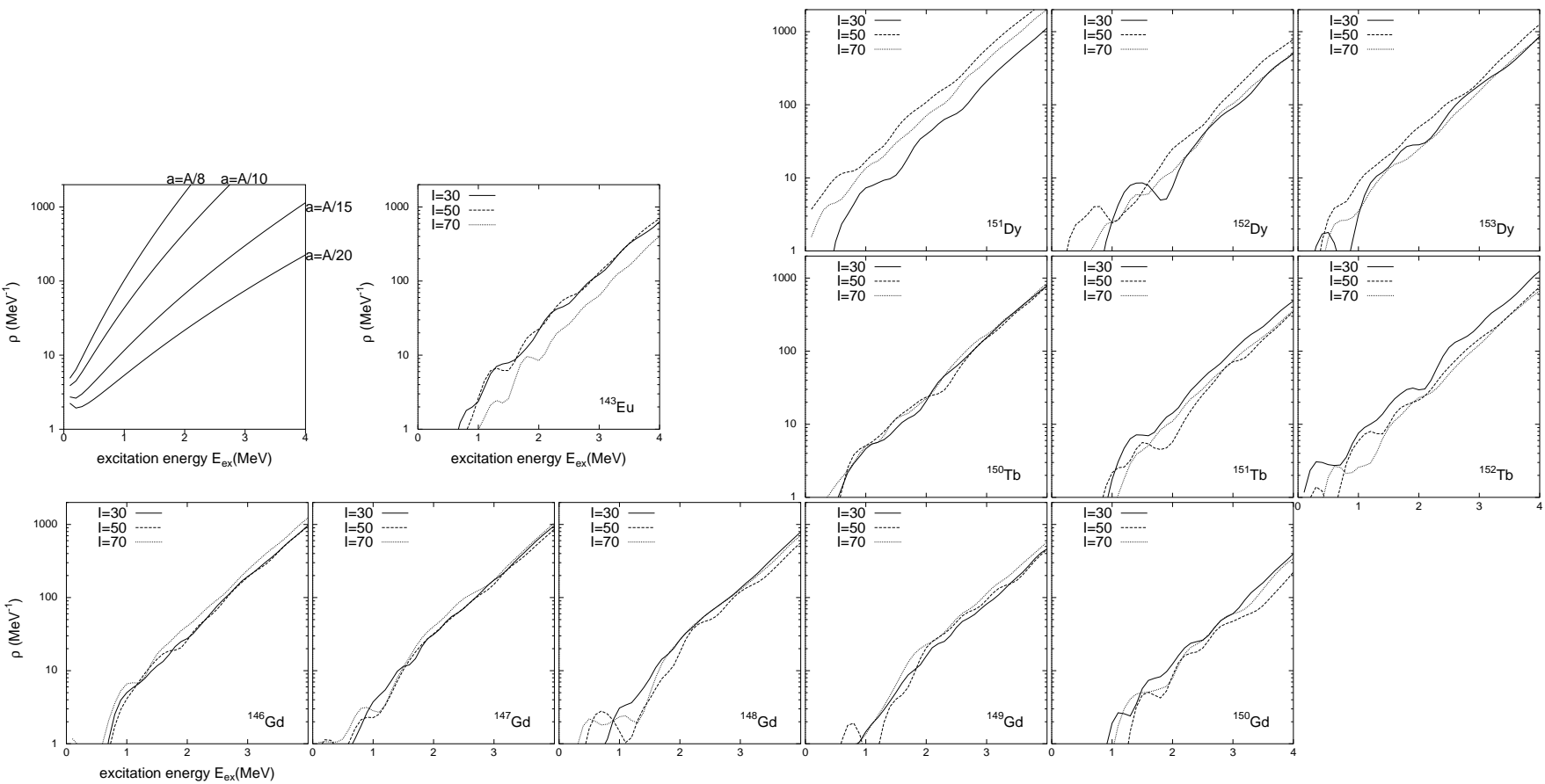}
}
\caption{\label{levdens} The level density of superdeformed states in
$A\sim 150$ nuclei as a function of the excitation energy measured
from the yrast line, averaged over both parities and
signature states at spin $I$ and  $I+1$ for 
$I=30,50,70\hbar$($I=\frac{59}{2},\frac{99}{2},\frac{139}{2}\hbar$ for 
odd nuclei). The Fermi gas formula are displayed
in the left-top panel with several $a$ parameters where $A=150$. 
} 
\end{figure}
\begin{figure}[tbp]
\centerline{
\epsfysize=22cm\epsffile{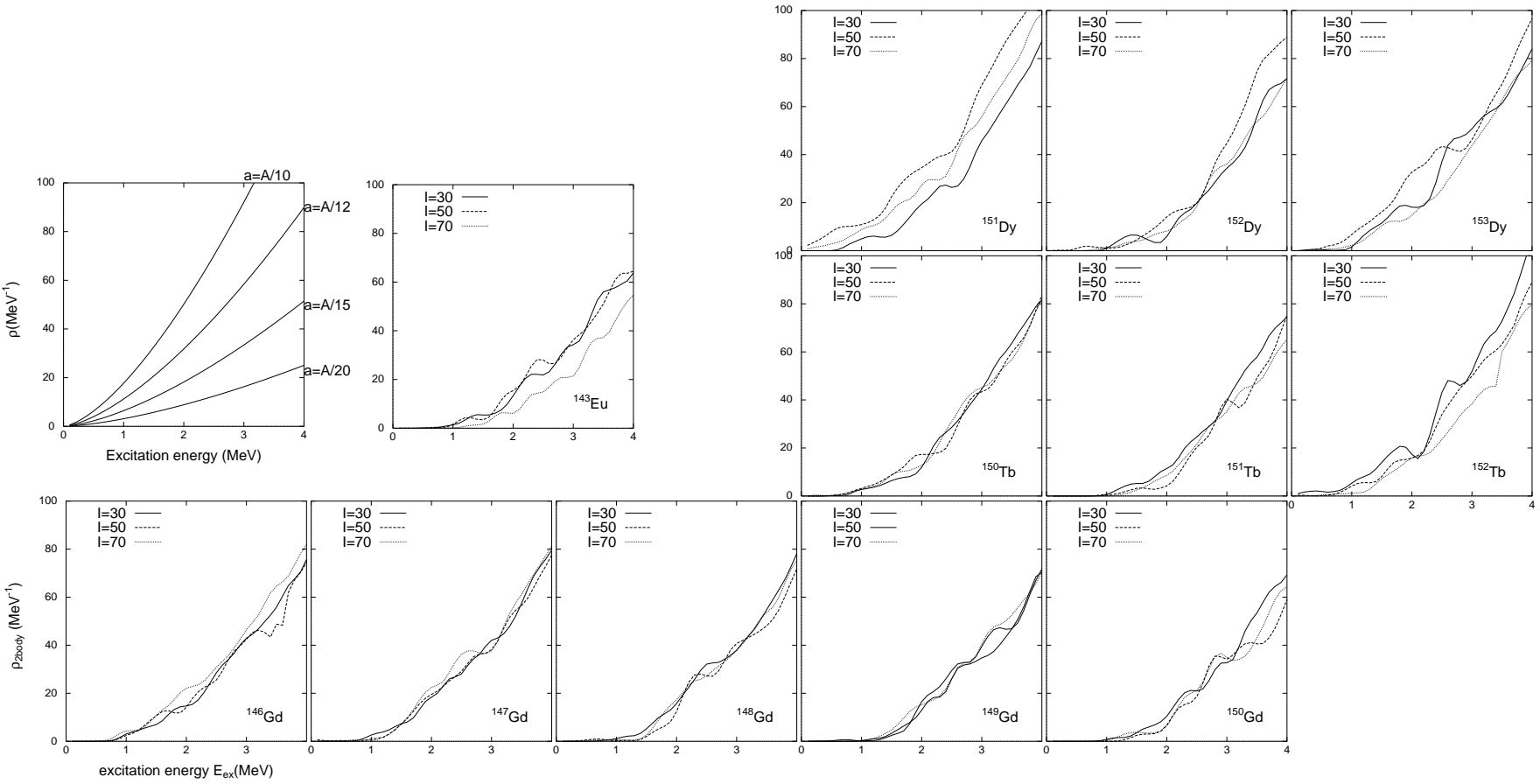}
}
\caption{\label{tlevdens}The two-body level density for superdeformed
states in $A\sim 150$ nuclei as a function of the excitation energy
measured from the yrast line, averaged over 
both parities and signature states at spin $I$ and  $I\! + \! 1$ for 
$I=30,50,70\hbar$($I=\frac{59}{2},\frac{99}{2},\frac{139}{2}\hbar$ for 
odd nuclei). The Fermi gas formula are displayed
in the left-top panel with several $a$ parameters where $A=150$. 
} 
\end{figure}
On the other hand,we find from similar decomposition of  the
two-body level density  
that the major contribution to $\rho_{\rm 2body}$ comes from the
component representing the interaction of 1p-1h to 1p-1h and 2p-2h to
2p-2h states.
The subcomponent associated with
the interaction among  $n$p$n$h states depends only linearly on energy as 
$\rho_{\rm 2body, {\it n}p{\it n}h} \propto (E-n\Delta)^{1}$. Thus the
gap is less effective for $\rho_{\rm 2body}$ than for $\rho_{\rm tot}$, explaining
the result shown in Fig.~\ref{ferns} and Table.~\ref{tabdns} 
that the level density parameter $a_{\rm tot}$ fitting
the total level density $\rho$ is systematically 
smaller than $a_{\rm 2body}$ of the two-body density $\rho_{\rm 2body}$.

The calculated level densities of superdeformed states  in
$A \sim 150$ region are plotted 
in Fig.~\ref{levdens} and 
\ref{tlevdens}.
Sizable change  in the level densities depending on neutron and proton 
number $N$ and $Z$ is
remarkable. For example, there is variation of more than factor
2 in  the total density
 $\rho(E)$. The
fitted level density parameters $a_{\rm tot}$ and $a_{\rm 2body}$ vary
50\% as shown in  Table.~\ref{tabdns}.
The variation of the level densities should be originated from
the changes in the single-particle  structure  in different
superdeformed nuclei, which are associated with the
changes in the Fermi surface and the deformation.
In order to demonstrate the correlation between the level densities
and the single-particle structure,we compare in  Table.~\ref{tabdns}, 
systematics of the fitted level density parameters with that of  the
shell correction energy $E_{\rm shell} = -(E^{\rm micro}-E^{\rm smooth})$
of the reference configuration, which is a
measure of 
the shell effects in the single-particle spectrum.
The correlation is clear in Table.~\ref{tabdns};
the larger is the shell correction energy, the smaller
become the fitted level density parameters 
$a_{\rm tot}$ and $a_{\rm 2body}$.
Taking Gd isotopes as an example, 
the level densities  increase as the neutron number decreases
from 86 to 82. Correspondingly the shell correction energy decreases for
lighter isotopes. This can be related to both the neutron and the proton 
single-particle routhian 
structure. In $^{150}$Gd ,the routhian spectra are not very different
from Fig.~\ref{gdsp}. The neutron and the proton spectra has a large
gap at $N=86$ and $Z=64$ Fermi surface ,respectively. For lighter Gd
isotopes ,the Fermi surface is situated at one to several orbits below 
the $N=86$ gap,where the single-neutron level density is rather high.
Furthermore the proton gap at $Z=64$ Fermi surface decreases as the
quadrupole deformation decrease(from $\epsilon_2= 0.556 $ at
$^{150}$Gd to  
$\epsilon_2=0.510  $ at $^{146}$Gd). Both effects cause the increase of 
level density and the decrease of the shell correction energy for
lighter Gd isotopes .
Another example is $^{151}$Dy ,in which the level density is the
largest among the calculated nuclei. This is because
there is no big gap in the neutron spectrum at $N=85$ Fermi
surface, and 
the proton gap at Z=66 is not very large (The proton
gap for $^{151}$Dy is smaller than that in 
 Fig.~\ref{gdsp} because of smaller quadrupole deformation).
In a few nuclei, the calculated level densities show
dependence on spin of about factor 2. The
significant spin dependence is also 
attributed to the changes in the shell gap. For example,
in $^{151}$Dy, the Z=66 proton gap just above the $6_4$ orbit
decreases with increasing rotational frequency up to $\omega \approx
0.5 $MeV/$\hbar$  
because of the signature splitting of $6_3, 6_4$ orbits (This
behavior is seen in Fig.~\ref{gdsp}).

\section{Onset of rotational damping}
\subsection{Results}
\begin{figure}[tbp]
\centerline{
\epsfxsize=15cm\epsffile{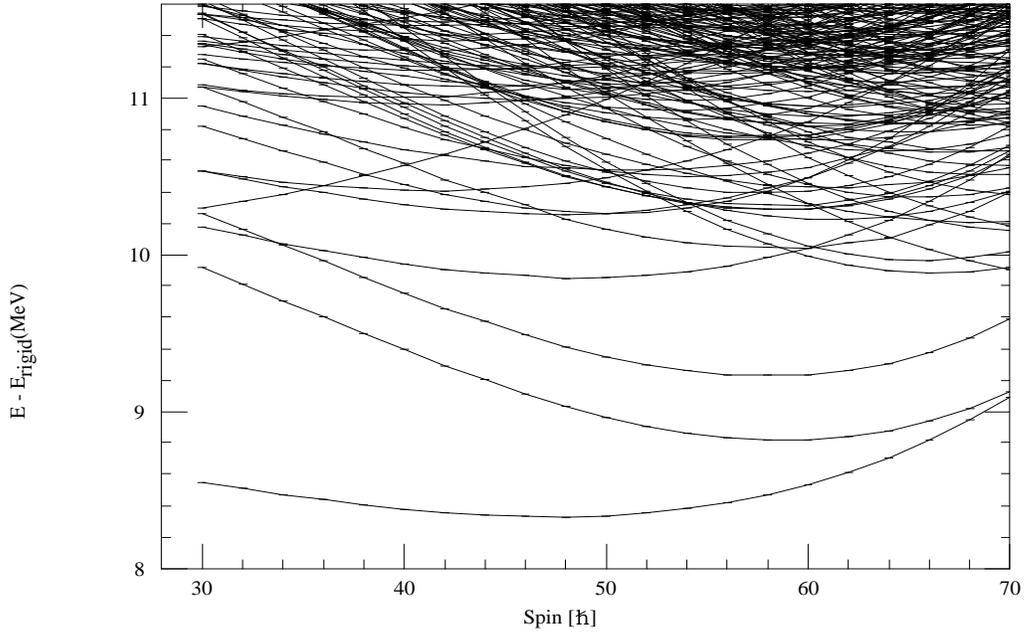}
}
\centerline{
\epsfxsize=15cm\epsffile{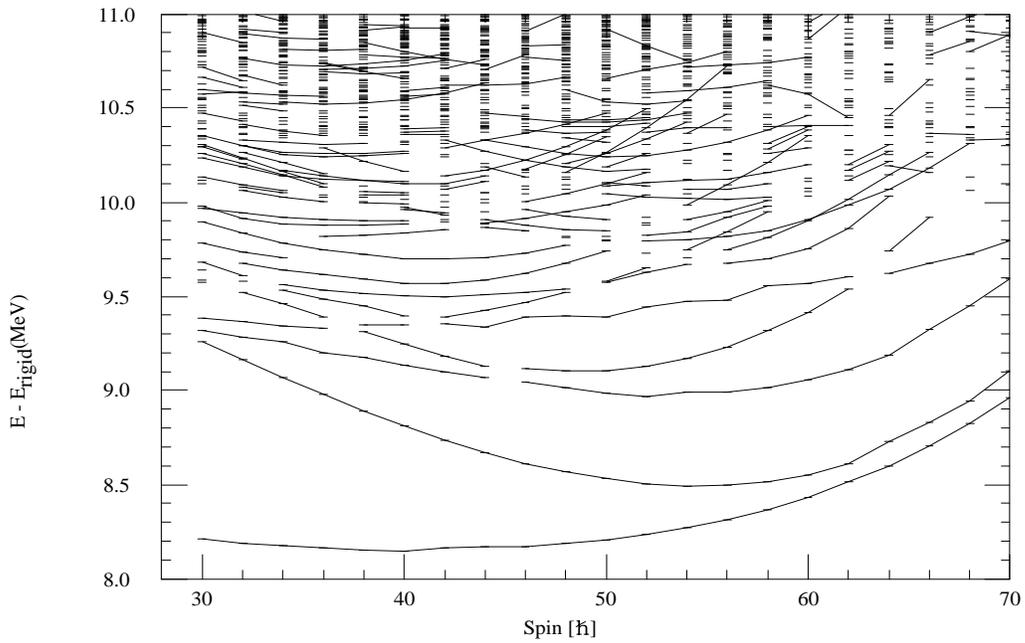}
}
\caption{\label{flow}The calculated energy levels plotted
with little horizontal bars for (0,+)states in $^{152}$Dy. An rigid-rotor
rotational energy 
$E_{\rm rigid} = I(I+1)/190$ (MeV) is subtracted.
Solid lines connecting the energy levels represent strong E2
transitions with normalized strength larger than $ \protect
\sqrt{1/2}=0.707$. 
The upper panel shows unperturbed levels with no residual
interaction.
The lower panel shows the results with the residual interaction.
}
\end{figure}

\hskip 12pt
Every unperturbed many-particle and many-hole configuration created
upon the cranked single-particle orbits would form a superdeformed
rotational band  
if there were no configuration mixing.
The unperturbed rotational bands are displayed 
in the 
upper panel of Fig.~\ref{flow} for the 
signature and parity  quantum number  $(0,+)$ in $^{152}$Dy.
 The lowest band has the configuration
filling the single-particle orbits  up to the Fermi level
at N=86 and Z=66 (See Fig.~\ref{gdsp}), which corresponds to the
observed yrast superdeformed band. The second lowest band is
a proton 1p-1h configuration with the hole at the $6_4$ orbit and
the particle at [413]$\frac{5}{2}$. Other unperturbed bands displayed
in the 
figure have  mostly 1p-1h and 2p-2h configurations.

With the residual two-body interaction included, the
unperturbed configurations are mixed to form energy eigenstates. The
resultant energy levels are plotted with
little horizontal bars in the lower panel of Fig.~\ref{flow} for (0,+)
states 
in $^{152}$Dy. 
The configuration mixing due to the residual interaction is strongly
dependent on the excitation energy. 
An eigenstate near the yrast consists mostly of a single dominant
configuration, while 
a state at high excitation energy consists of many unperturbed 
configurations. Correspondingly, the E2  decay from a near-yrast level
at $I$ feeds to 
a level at $I\! -\! 2$ containing the same dominant component.
The associated E2 strength exhausts most of the total strength for the 
collective rotation.
The solid lines
connecting the levels in Fig.~\ref{flow} correspond to such strong
E2 transitions. On the other hand, 
the E2 strength for higher excited levels is fragmented over
small components decaying to many final states.
A sequence of the strong collective E2 transitions (solid lines)
indicates a rotational band structure. The band structure is
dominant near the yrast line while it tends to disappear as
the excitation energy increases, indicating the onset of the rotational
damping.

\begin{figure}[tbp]
\centerline{
\epsfysize=21.5cm\epsffile{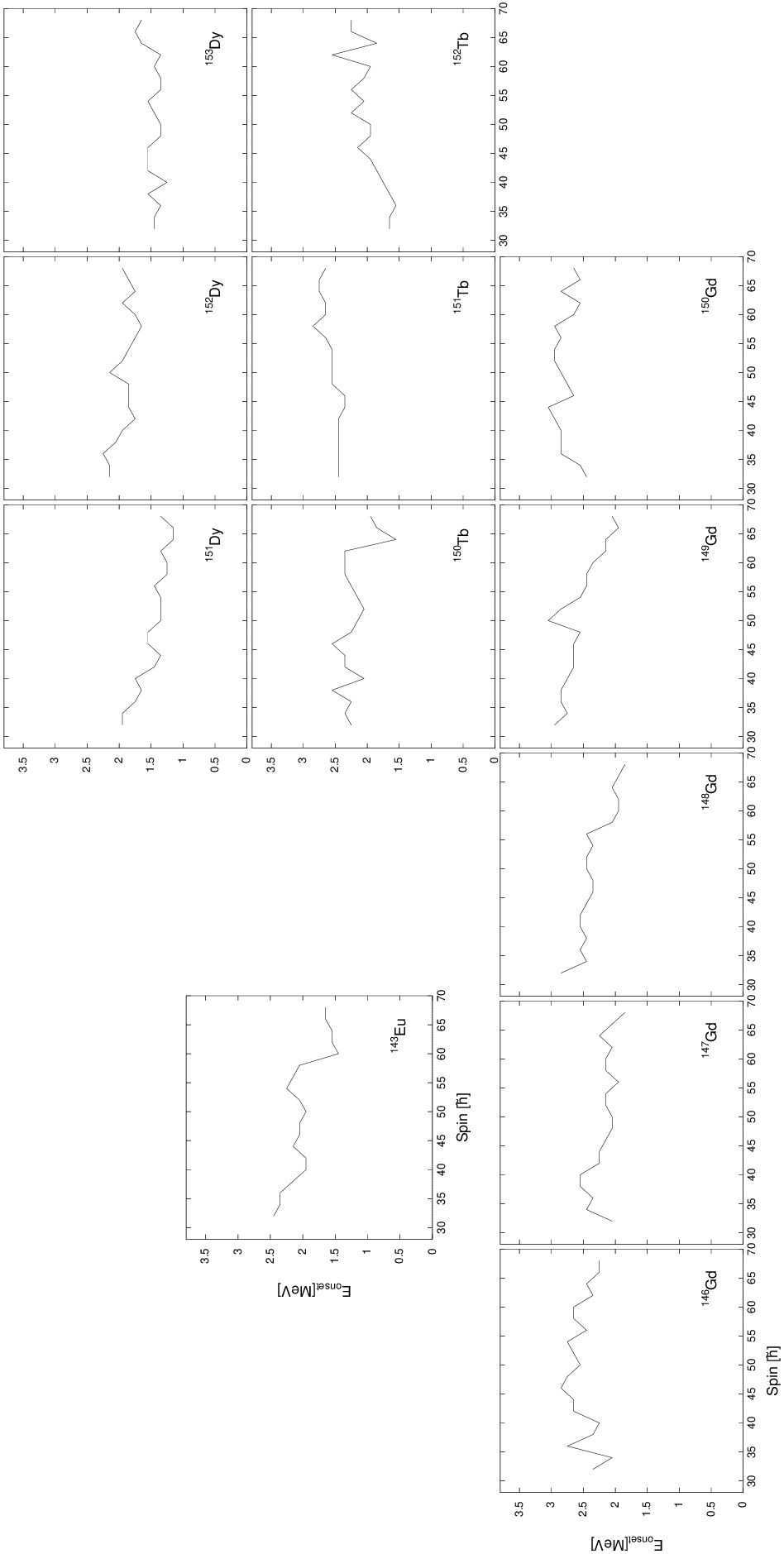}
}
\caption{\label{onset}Onset excitation energy of rotational damping
$E_{\rm onset}$ measured from the yrast line for each  superdeformed
nuclei,as a function of spin.
}
\end{figure}

In order to make a systematic analysis of the
onset of rotational damping, let us discuss  the excitation energy
where the 
damping sets in. It is useful to define the branching
number\cite{Matsuo93} 
\begin{equation}
n_{\rm branch}(\alpha)\equiv\left(\sum_\beta w_{\alpha
\rightarrow \beta} ^2\right)^{-1}
\end{equation}
associated with the E2 strength for decays 
from a level $\alpha$. 
Here  $w_{\alpha\rightarrow \beta}$ is the
normalized E2 strength, Eq.~(\ref{eqbranch}), from a level 
$\alpha$ at $I$ to a
level $\beta$ at $I\! -\! 2$. This quantity measures effectively 
the number of E2 decay branches. The onset of the damping can be
defined by the condition that the E2 decays have more than
two branches or $n_{\rm branch} >2 $ \cite{Matsuo93,Matsuo96}.
\begin{figure}[tbp]
\centerline{
\epsfysize=10cm\epsffile{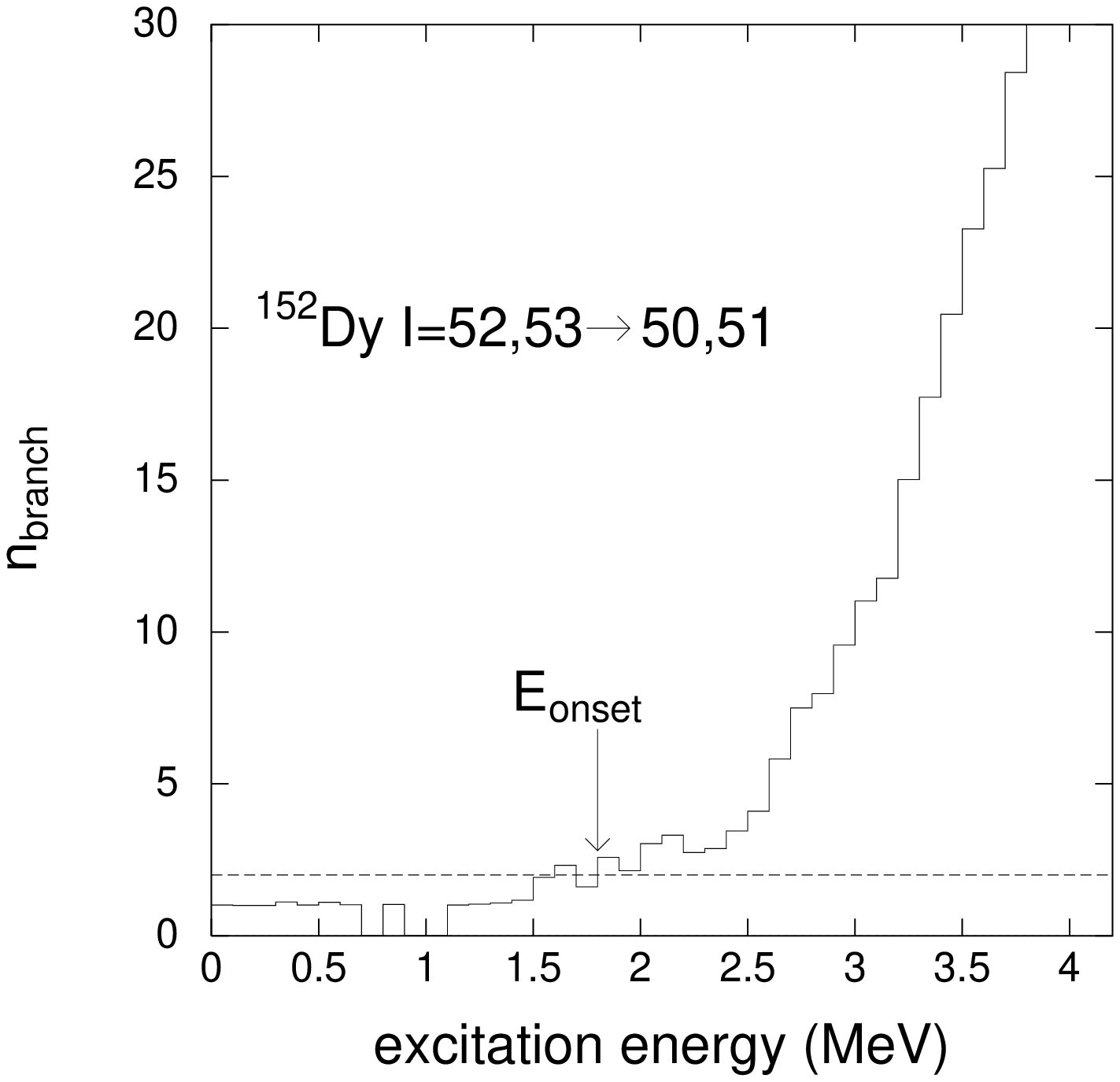}
}
\caption{\label{branch}The branching number $n_{\rm
branch}$ as a function of  excitation energy 
above yrast line for  $^{152}$Dy, averaged
over $I=52^{\pm}\rightarrow 50^{\pm},I=53^{\pm}\rightarrow
51^{\pm}$. The horizontal lines 
shows $n_{\rm branch}\! =\! 2$ used to define the onset of
 rotational damping. 
}
\end{figure}
An example of the calculated $n_{\rm branch}$ is plotted in
Fig.~\ref{branch} 
as a function of the excitation energy measured from the yrast line. 
The branching
number increases with excitation energy, and the
onset energy of the damping $E_{\rm onset}$ where $n_{\rm branch}$
exceeds 2 is 
given as 1.8 MeV in this example.
The calculated onset energy is shown in Fig.~\ref{onset} as a function of
spin for superdeformed nuclei in $A\sim 150$ region.

Because of the onset of the rotational damping, there exist
only limited number of rotational bands  which show up as 
 the regular sequences of levels connected with strong 
collective E2 transitions. The number of
rotational 
bands thus gives a quantitative measure of the onset of the damping.
Noticeably, there exist many  ``rotational bands with short length''
in which strong E2 transitions continue only for two or three steps.
In defining the number of rotational bands, we count the bands
with length more than two. This definition corresponds to the
experimental definition of the effective number of paths
\cite{Dossing95} which  
is extracted from the spectral fluctuation at the first ridge 
of the $E_{\gamma_1}  \times  E_{\gamma_2}$ spectra, which are formed
by two consecutive  
E2 transitions along rotational bands. 
In practice,
we require that along a rotational band the strong collective
E2 transitions continue at least two steps with the criterion
$n_{\rm branch} < 2$.
Counting those levels satisfying this condition 
for all parities and signatures, say
$I^\pi=40^+,40^-,41^+,41^-$, we define the number of bands $N_{\rm
band}$ at 
a representative spin, $40\hbar$ in this example. 
\begin{figure}[tbp]
\centerline{
\epsfysize=22cm\epsffile{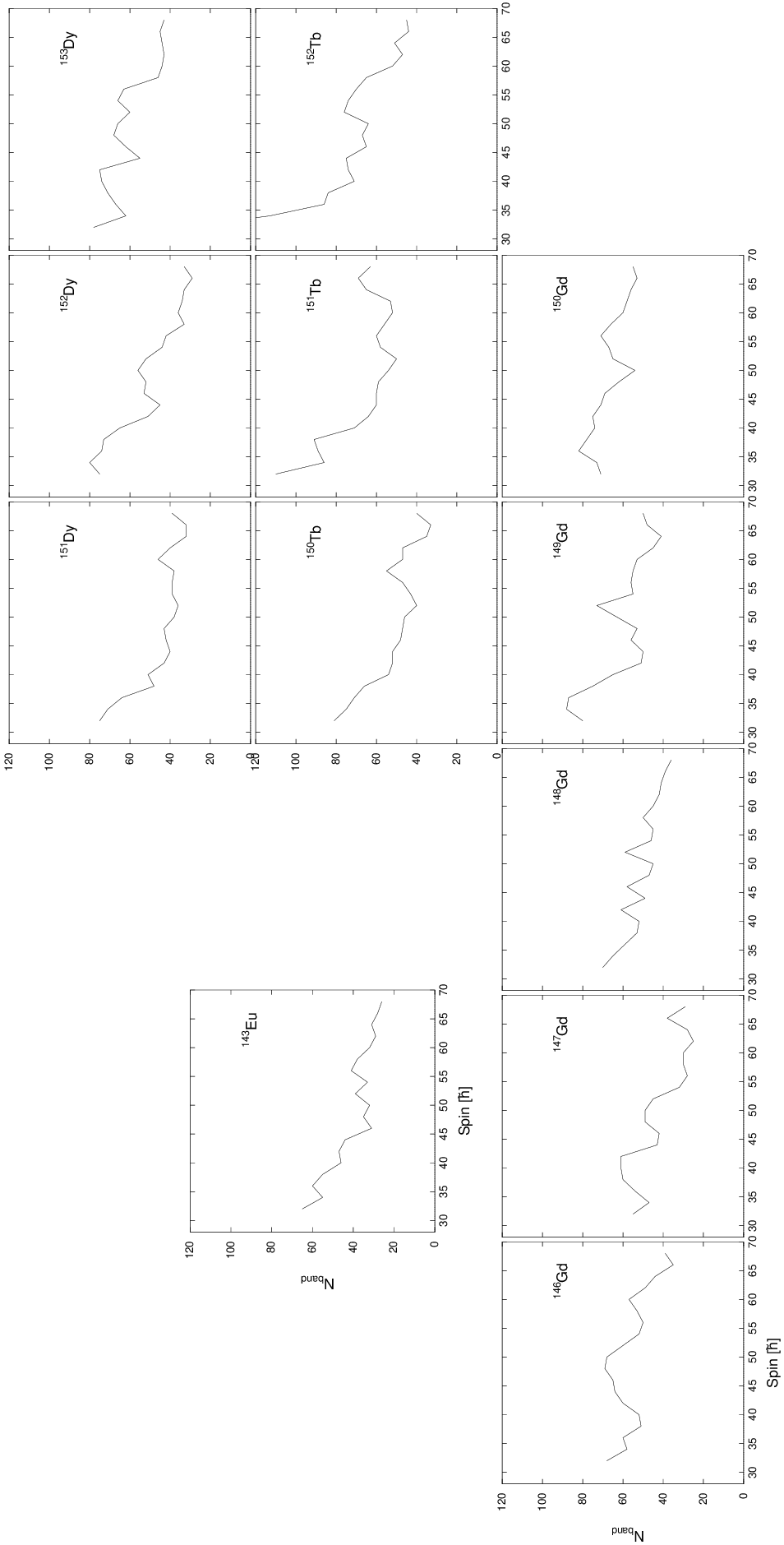}
}
\caption{\label{numbands}The calculated number of superdeformed bands 
$N_{\rm band}$ (see text for definition) as a function
of spin for each superdeformed nucleus.
}
\end{figure}

The result is plotted
in Fig.~\ref{numbands}. The fluctuations in the calculated $N_{\rm
band}$ depending on spin 
arise from the sharp boundary $n_{\rm branch} < 2$ defining rotational
bands, and it may rather represent  ambiguity of the definition
of $N_{\rm band}$. In order to see the dependence of the species,it
is preferable to take averages of the onset 
energy $E_{\rm onset}$ and the number of bands $N_{\rm band}$ over
different spins. 
\begin{table}
\begin{center}
\begin{tabular}{|c|c|c|}
\hline
& $E_{\rm onset}$[MeV] & $N_{\rm band}$ \\
\hline
$^{143}$Eu& 2.0 &         38      \\
$^{146}$Gd& 2.7 &         60      \\
$^{147}$Gd& 2.1 &         41      \\
$^{148}$Gd& 2.3 &         51      \\
$^{149}$Gd& 2.6 &         57      \\
$^{150}$Gd& 2.9 &         66      \\
$^{150}$Tb& 2.3 &         48      \\
$^{151}$Tb& 2.6 &         57      \\
$^{152}$Tb& 1.7 &         67      \\
$^{151}$Dy& 1.4 &         40      \\
$^{152}$Dy& 1.8 &         46      \\
$^{153}$Dy& 1.5 &         60      \\
\hline
\end{tabular}
\caption{\label{tabdmp} The calculated onset excitation 
energy $E_{\rm onset}$ 
of rotational damping measured from the yrast line, and 
the calculated number of superdeformed bands $N_{\rm band}$,
averaged over $I=40\hbar \sim 60\hbar $. 
}\end{center}
\end{table}
Table.~\ref{tabdmp} summarizes 
the average $E_{\rm onset}$ and $N_{\rm band}$ where the averages are
taken over 
spin range $I=40\sim 60\hbar$, which approximately correspond to 
spins where the  superdeformed states are expected to dominate
the  yrast region. 

It should be emphasized that, although the onset energy thus defined
tells approximately where the rotational damping sets in,
the transition from the
rotational band structure to
the rotational damping is not very sharp as function
of excitation energy, but it develops 
rather  gradually. As shown in Fig.~\ref{flow}, there are
numerous ''short bands'' which are surrounded by the excited levels
with damped rotational transitions. Because of the presence of
such scars of rotational bands\cite{Broglia96}, the onset energy
$E_{\rm onset}$ should be regarded as one of the measures defining
the transition. In fact, the scars still exist at 1 MeV above the 
onset energy. 

\subsection{Discussion}
\hskip 12pt
The calculated onset energy $E_{\rm onset}$ for the superdeformed nuclei
is about $1.5\sim 3.0$ MeV above yrast line, depending on species 
(Fig.~\ref{onset} and Table.~\ref{tabdmp}). It is significantly higher
than the onset energy 
$E_{\rm onset}\sim 0.8$ MeV in normal deformed rare-earth
nuclei\cite{Matsuo93,Matsuo96}.
The calculated number of bands $N_{\rm band}\! \sim\! 40 -\! 70$, is
also larger 
than the corresponding number $N_{\rm band} \sim 30$ in normally
deformed nucleus
$^{168}$Yb, while $N_{\rm band}$ in a few superdeformed nuclei such as
$^{143}$Eu and $^{151}$Dy is comparable with that in normal deformed
nuclei.  

Another characteristic feature seen from
Fig.~\ref{onset},\ref{numbands} and 
Table.~\ref{tabdmp} is variation of  the onset energy and the number of
bands in different superdeformed nuclei.
For example, the onset energy   becomes the
highest(about 3.0 MeV) and the number of bands is also large(about 70) in
$^{150}$Gd,  while the same quantities become
a factor 1/2 small in  $^{143}$Eu (about 2.0 MeV and 40) 
and $^{151}$Dy (about 1.4 MeV and 40). One can also trace that
$E_{\rm onset}$ and $N_{\rm band}$ decrease as the neutron number
decreases 
in Gd isotopes. 

Both the difference from the normal deformed nuclei and the variation
among the superdeformed nuclei are explained in terms of the
characteristics 
of the level densities discussed in Sec.~\ref{leveldns}.
In this connection, it is useful to recall the argument in
Ref.\cite{Lauritzen86} that
the damping of rotational motion  takes
place if the spreading of unperturbed $n$p-$n$h
configurations is caused by the two-body residual interaction, which
is represented by the condition $\Gamma_{\mu} > d_{\rm 2body}$. Here 
$\Gamma_{\mu}=2\pi v^2 \rho_{\rm 2body}$ is the
spreading width of $n$p-$n$h configurations 
with $v$ being the average size of  
matrix elements of the residual interaction, and $\rho_{\rm 2body}$
being the two-body level density. 
The spreading is 
meaningful only when it is larger than the spacing between
interacting states
$d_{\rm 2body}=\rho_{\rm 2body}^{-1}$. The above condition reads 
\begin{equation}
\rho_{\rm 2body} >1/\sqrt{2\pi}v
\label{criteria}
\end{equation}
implying that the onset of the rotational damping is governed by the
ratio between the two-body level spacing $d_{\rm 2body}=\rho_{\rm
2body}^{-1}$ and the 
two-body 
interaction strength. As 
$\rho_{\rm 2body}(E)$ increases with excitation energy,
the rotational damping sets in at the
excitation energy where the condition Eq.~(\ref{criteria}) is satisfied.
It is now to be noticed that the two-body level
density in superdeformed nuclei is much lower than in normally
deformed nuclei as shown in Fig.~\ref{ferns}.
Assuming that the matrix
elements of residual two-body interaction does not depend
on deformation,  the onset energy $E_{\rm onset}$ satisfying the
condition Eq.~(\ref{criteria}) must be much larger 
in superdeformed nuclei than in normally nuclei.
The variation of $E_{\rm onset}$ among the superdeformed nuclei
can be explained along the same argument. It is  seen
from combination of Table.~\ref{tabdns} and Table.~\ref{tabdmp} that
the variation 
of $E_{\rm onset}$ is correlated with that of the fitted level density
parameters 
and the shell correction energy. For example, the onset
energy $E_{\rm onset}$,
 the fitted level
density parameters and the shell correction energy become the
largest in $^{150}$Gd. In this nucleus, the two-body level density
becomes very small due to the shell gap at N=86 and Z=64 Fermi
surfaces.

To be more quantitative, let us adopt the Fermi gas expression
Eq.~(\ref{twobody}) for the two-body level density. Then the onset
energy is estimated from Eq.~(\ref{criteria}) as 
\begin{equation}
\label{eqon}
E_{\rm onset}=  7.11 \bar{v}^{-2/3}a^{-5/3}.
\end{equation}
Choosing 
$v=25 $ keV and $a=A/10 (A=170)$, the onset energy
in this estimate
 $E_{\rm onset}\! =\! 0.74 $ MeV agrees with the microscopic
calculation  for normally deformed nuclei\cite{Matsuo96}. With the
same interaction  
strength, but with smaller level density parameter
$a=A/15\sim A/20 (A=150)$ corresponding to the superdeformed states,
the estimate reads  $E_{\rm onset}\! =\! 1.8\sim 2.9$ MeV, which
roughly agrees with the microscopic results in Table.~\ref{tabdmp}.

Let us now discuss the number of bands.
The microscopic results are about $40\sim70$, which are
about factor 1-2 larger than in normally deformed nuclei.
The variation of $N_{\rm band}$ among the superdeformed nuclei
is correlated to that of the level densities as seen from
Table.~\ref{tabdns} 
and Table.~\ref{tabdmp}; the larger the number of bands is, the
smaller are the level density parameters.
For qualitative understanding of these features, let us 
utilize two schematic models. The first one is along the
same line with the above discussion for $E_{\rm onset}$. 
The  number of bands in this model may be estimated 
by counting the levels up to the onset energy given by Eq.~(\ref{eqon})
provided that the boundary is sharp(while,in fact, 
this assumption is not necessarily correct as Fig.~\ref{flow} indicates).
Then the number of bands
is given by
\begin{eqnarray}
\label{analytic}
N_{\rm band}&=&4\int_0^{E_{\rm onset}}\rho(E)dE \nonumber \\ 
&=&\frac{\sqrt{\pi}}{12}\int^{x_0} x^{-5/4}\exp{2\sqrt{x}}dx,
\end{eqnarray}
\[
x_0=7.11(a\bar{v})^{-2/3}.
\]
Here the Fermi gas expression Eq.~(\ref{fermtot}) for the total level
density 
$\rho(E)$ is used. Multiplier 4 is needed to count different parity
and signature states.
The number of bands is an steeply
increasing function of $(av)^{-1}$;\\
$N_{\rm band}\!=26 ,55,100 ,170 ,270,400  $ for
$(av)^{-1}\! =2,3,4,5,6,7$. 
The choice of the interaction strength $v\! =\! 25 $ keV gives
$N_{\rm band}\! =\! 34 $ for a normally deformed nucleus with $A\! =\!
170$ in agreement with 
the microscopic calculation. With small level density
parameters corresponding to the superdeformed nuclei, 
the estimate produces 
$N_{\rm band}=\! 101\sim\! 201$ for $a=\! A/15 \sim\! A/20$. This
indicates the 
significant increase of the number of superdeformed bands and its
sensitivity to the variation of  
level density. However
this model apparently overestimates the microscopic results
in Table.~\ref{tabdmp}.
The overestimate in $N_{\rm band}$ can be traced back 
to the fact already revealed that 
the density parameter
$a$ fitting the total level density $\rho(E)$ 
is significantly smaller than
that for $\rho_{\rm 2body}(E)$. As is discussed in Sec.~\ref{leveldns}, 
the level densities affected by the presence of the shell
gap in the single particle spectrum 
cannot be simulated  simply by a single value of the
the level density parameter.
In order to  treat the shell gap explicitly, let us next consider
the other model, which simplifies the situation by assuming
that the single-particle spectrum is uniform with density
$\frac{1}{2}g_0$ 
(related to the level density parameter $a\! =\! \frac{\pi^2}{6} g_0$) 
except the presence of the gaps
with interval $\Delta$ at the Fermi surface.
Considering  the limit of large $\Delta$, the lowest excited levels are
dominated only by 1p-1h excitations. In this limit, the total 
level density $\rho(E)$ is significantly reduced, and 
is expressed as
\begin{equation}
\rho(E) =\rho_{\rm 2body}(E) = \frac{1}{2}g_0^2  (E-\Delta) \hskip 10mm 
(E>\Delta) 
\end{equation}	
The two-body level density is the same as $\rho(E)$ since all
the 1p-1h states can interact through the two-body force. 
With $v\! =\! 25$ keV and $a\! =\! A/10$ corresponding to the average
single-particle density\cite{Bohr-Mottelson69I},
the onset energy and the number of bands
are estimated as
\begin{eqnarray}
E_{\rm onset}&=&\frac{4\sqrt{2}}{\sqrt{\pi}vg_o^2}+\Delta 
=  1.53 {\rm MeV}  + \Delta ,\\
N_{\rm band}&=& \frac{8}{\pi v^2 g_0^2} = 49 .
\end{eqnarray}
The onset energy $E_{\rm onset}$ increases with increase
of the single-particle gap $\Delta$ while the estimated number of bands
reaches to a constant independent of $\Delta$,which is  not very large
comparing 
with an estimate  
$N_{\rm band}\! =\! 44$ corresponding to $\Delta=0$($a=A/10,A=150$ in
Eq.~(\ref{analytic})). The two schematic models discussed
above simplify the realistic features of the single-particle spectra
for superdeformed potential.
 The calculated level densities
presented in Sec.~\ref{leveldns} seem to lie in-between the two models.
 
It should be reminded that the onset of rotational damping does not take
place very sharply at the onset energy, but proceeds quite irregularly
as function of the excitation energy, as is discussed in the
preceding subsection  and demonstrated in Fig.~\ref{flow}. This is because
there are local fluctuation of the unperturbed levels and the matrix elements
of the residual interactions depending on the configurations of the
many-particle many-hole states. 
A little deviation of $E_{\rm onset}$ and $N_{\rm band}$ from the
systematics expected from the $N,Z$-dependence of the
shell structure effect in  the single-particle levels may be due to
such fluctuations.

We noticed,  
in Fig.~\ref{numbands}, systematic spin dependence that the
number of bands increases with decreasing  spin below $40\hbar$.
This spin dependence is not correlated with the level density since
the calculated level densities in Fig.~\ref{levdens} and ~\ref{tlevdens}
have no systematic spin dependence 
which can explain the increase of $N_{\rm band}$.
This may indicate that the onset of rotational damping is not
controlled by the configuration mixing alone.
It is should be noticed 
in this connection that the alignment effect is also important to
cause the
rotational damping, while it is not very large for the
low spin region in superdeformed nuclei. 
The dispersion of rotational frequency $2\Delta \omega$,which measures 
the alignment effect and represents 
how different rotational bands respond to the spin change of $\Delta I=2$,
is about 40 keV at $I=30$ on the basis of the estimate of
Ref.~\cite{Lauritzen86}. 
This is not sufficiently large than
the scale of the residual interaction effects $\sim 50$ keV
($d_{\rm 2body}\sim \sqrt{2\pi}v$  at the onset energy),
the rotational correlation persist more than expected from
the argument based on the level densities.
It is to be reminded that possible persistence of the rotational band
structure in the region of strong configuration mixing is pointed out
in Ref.~\cite{Mottelson92}.

\subsection{Comparison with experiments}
\hskip 12pt
Recent experiments identify several superdeformed bands in
a nucleus in $A \sim 150$ region\cite{Dagnall94}. 
Our
microscopic calculation predicts 
at least more than ten superdeformed rotational bands even if ''the
short rotational 
bands'' are excluded (there are three other figures 
in addition to the lower panel of Fig.~\ref{flow} for
$^{152}$Dy). It implies that 
the current spectroscopic 
studies  of discrete gamma-ray peaks reaches far below the onset
of rotational damping in superdeformed nuclei. In other words, the
present calculation  suggests possibility
to find further number of superdeformed bands in a nucleus
in future experiments.

Available experimental data concerning the 
the onset of rotational damping in superdeformed nuclei
come from analysis of quasi-continuum part of 
double-coincident  $E_{\gamma_1}  \times  E_{\gamma_2}$ spectra
although the number of data 
is very limited so far.
An indirect information comes from an simulation of the
quasi-continuum gamma-rays in superdeformed $^{152}$Dy 
\cite{Schiffer91}, which  points to an estimated value
$E_{\rm onset}$ around $2.5\sim 3.0$ MeV from the observed
intensity of the ridge part of the $E_{\gamma_1}  \times
E_{\gamma_2}$  spectra \cite{Twin88}. This analysis agrees with the
present microscopic calculation in the qualitative feature that
the onset of damping takes place at  much higher excitation energy
than in normal deformed nuclei.

More direct information comes from the fluctuation analysis method
which extracts the effective number of paths of decays
from the quasi-continuum part  of  $E_{\gamma_1}  \times
E_{\gamma_2}$ spectra 
\cite{Herskind92,Dossing95}.  
When this method
is applied to the first ridge of the spectra located at
$E_{\gamma_1}-E_{\gamma_2}=\pm 4/\cal{I}$ lines with $\cal{I}$ being
the moment of inertia of 
superdeformed bands, the effective number of paths counts the
number of two consecutive collective E2 transitions, $I\! +\! 2
\rightarrow\! I\! \rightarrow I\! -\! 2$.
Thus the effective number of paths at the first ridge approximately
corresponds to the number of bands which is defined in Sec.~\ref{form}.
The data for superdeformed nuclei is available only 
in $^{143}$Eu \cite{Leoni95}.
\begin{figure}[tbp]
\centerline{
\epsfysize=10cm\epsffile{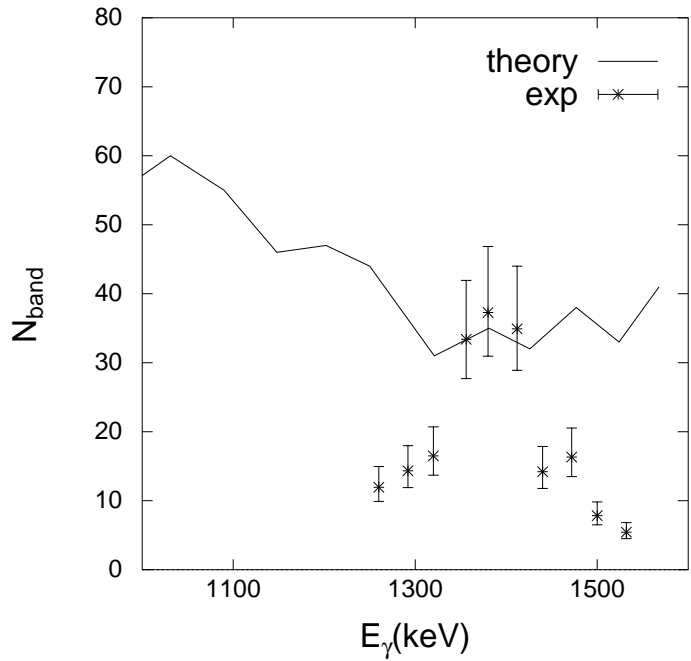}
}
\caption{\label{exp}The experimental effective number of decay paths 
\protect \cite{Leoni95} and the calculated number of superdefomed bands
$N_{band}$ as a function of the transition gamma-ray energy.
}
\end{figure}
Comparison between the theory and the experimental data is done
in Fig.~\ref{exp}, in which our results is plotted as a function of the
average gamma-ray energy.
The data points are limited in the region $E_{\gamma} \sim 1250 -
1550$ keV 
where the quasi-continuum ridge is observed.
The calculated number of bands agrees with 
the experimental effective number of paths 
at around $E_{\gamma} \sim 1400 $keV.  

The effective number of paths becomes
smaller around high and low ends of the measured $E_{\gamma}$ range
than the 
theoretical number of bands. This may suggest 
some other mechanisms that destroy rotational
band structures in addition to the rotational damping. 
For example, for the region with $E_{\gamma} \gesim 1500$ keV
corresponding to the very high spins $I \gesim 55\hbar$, the fission
decay may compete with the collective E2 gamma decay and  hinder the
E2 decay 
branches\cite{Schiffer91}. Furthermore ,feeding mechanism of the
superdeformed states may affect the effective number of paths.
The low gamma-ray energy $E_{\gamma}\lesim 1300$ keV, on the other 
hand, corresponds to spins $I \lesim 40\hbar$, at which the lowest
superdeformed states lie higher in energy than the yrast line
consisting 
of normally deformed states. Since the superdeformed states are
surrounded by numerous normal deformed states in this situation,
mixing between 
normal and superdeformed states may take place and cause
breaking of rotational band structures of the superdeformed states.
Inclusion of these effects has possibility to explain the data points, but
it is beyond the scope of the present paper.

\section{Conclusion}
\hskip 12pt
We analyzed the onset of the rotational damping in $A \sim 150$
superdeformed 
nuclei by means of the 
shell model diagonalization method comprised of the cranked Nilsson
mean-field and the residual two-body interaction.

The rotational
damping in superdeformed nuclei sets in at the excitation
energy of about $2\sim 3$ MeV above superdeformed yrast line. The
onset energy is much 
higher than the theoretical prediction in normally 
deformed rare-earth nuclei. 
The effective number of
rotational bands is calculated to be $40\sim 70$, which is slightly
larger 
than the corresponding number (about 30) in normally deformed nuclei.
Both the onset energy and the number of bands show the variation
depending on the superdeformed species even among $A \sim 150$ region.

All these features characteristic for the superdefomed nuclei
are originated from the shell structure in the single-particle spectrum 
associated with the superdeformed shape of the mean-field potential.
This was demonstrated explicitly by looking into the 
total level density as well as the two-body level density 
which is relevant to the configuration mixing via two-body forces.
The shell gap in the single-particle spectrum 
causes significant decrease in  the  level densities 
in comparison with those in the normally deformed
nuclei, and causes increase in the onset energy and the number of
bands. Since the shell structure at the Fermi
surface varies with  neutron and proton numbers and associated changes
in deformation, the level densities show
significant dependence on superdeformed species even among
$A\! \sim\! 150$ region. It causes the variation of the
onset of rotational damping.

The calculated results are consistent with existing data.
More systematic comparison with data is required, however, 
in order to check the predictions  such as the 
mass-number dependence, and also to clarify other mechanisms
which are not included in the present theory.

\section*{Acknowledgment}
\hskip 12pt
Discussions with S. Leoni, B. Herskind, T. D\o ssing, E. Vigezzi,
 R.A. Broglia and\\ 
K. Matsuyanagi are
greatly acknowledged. We also thank Y.R. Shimizu for discussion and
for providing us the liquid-drop code.
Numerical computation in this research was supported in part by Research Center for Nuclear
Physics(RCNP), Osaka University,  and the Institute for Chemical
and Physical Research (RIKEN).
A part of the numerical computations was carried out at RIKEN. 

\section*{Appendix}
\hskip 12pt
In order to calculate the two-body level density $\rho_{\rm
2body}(E)$, it is useful to define a related quantity
\begin{equation}
\label{define}
\rho^{(2)}(E_1,E_2)=\sum_{\mu\mu'}{}^{'}\delta(E_1-E_{\mu})
\delta(E_2-E_{\mu}') 
\end{equation}
where $\sum_{\mu\mu'}{}^{'}$ denotes the summation over pairs of 
$n$p-$n$h states which interact
 via two-body residual interaction. 
Eq.~(\ref{define})
can be rewritten as
\begin{eqnarray}
\rho^{(2)}(E_1,E_2)&=&\sum_\mu\delta(E_1-E_n)\sum_{\mu'(\mu)}{}^{'}\delta(E_2-E_{\mu'})
\\
&=&\sum_\mu\delta(E_1-E_\mu)\rho_{\rm 2body}(\mu,E_2) \\
&=&\rho(E_1)\rho_{\rm 2body}(E_1,E_2) 
\end{eqnarray}
Here the quantity 
$\rho_{\rm 2body}(\mu,E_2)$ representing density of states 
coupled with the state $\mu$ via the two-body interaction is
assumed to depend only on the energy $E_\mu$ of the state.
Then the two-body level density is given by
\begin{eqnarray}
\rho_{\rm 2body}(E)&=&\rho_{\rm 2body}(E,E) \nonumber \\
&=&\rho^{(2)}(E,E)/\rho(E) \ \ .
\end{eqnarray}
In numerical calculation, the delta functions in Eq.~(\ref{define})
is replaced by an Gaussian function  
used in the Strutinsky averaging. The level density
$\rho(E)=\sum_\mu \delta(E-E_\mu)$ is also calculated with the
same Gaussian function. The width of the Gaussian function 
used in the numerical calculation is  200 keV.

\end{document}